# Application of Machine Learning to Sporadic Experimental Data for Understanding Epitaxial Strain Relaxation

*Jin Young Oh[1], Dongwon Shin[2,*], and Woo Seok Choi[1,*]*

[1]*Department of Physics, Sungkyunkwan University, Suwon 16419, Republic of Korea*

[2]*Materials Science and Technology Division, Oak Ridge National Laboratory, Oak Ridge, TN 37831, USA*

Corresponding author e-mail : shind@ornl.gov, choiws@skku.edu

## ABSTRACT

Understanding epitaxial strain relaxation is one of the key challenges in functional thin films with strong structure-property relation. Herein, we employ an emerging data analytics approach to quantitatively evaluate the underlying relationships between critical thickness ($h_c$) of strain relaxation and various physical and chemical features, despite the sporadic experimental data points available. First, we have collected and refined reported $h_c$ of perovskite oxide thin film/substrate system to construct a consistent sub-dataset which captures a common trend among the varying experimental details. Then, we employ correlation analyses and feature engineering to find the most relevant feature set which include Poisson's ratio and lattice mismatch. With the insight offered by correlation analyses and feature engineering, machine learning (ML) models have been trained to deduce a decent accuracy, which has been further validated experimentally. The demonstrated framework is expected to be efficiently extended to the other classes of thin films in understanding $h_c$.

## 1. Introduction

Epitaxial strain and its relaxation mechanism in transition metal oxide thin films and heterostructures are critical for understanding and tailoring the strain-induced emergent functional properties.[1-3] The physical and chemical properties of perovskite oxide thin films are strongly affected by the microscopic lattice structure via sensitive structure-property relation, primarily altered by epitaxial strain and its relaxation. The in-plane lattice constant of the thin film follows that of the substrate up to a specific thickness, defined as the critical thickness ($h_c$) typically in the range of a few tens of nanometers because of the epitaxial strain imposed by the substrate.[4, 5] For films of thickness above $h_c$, epitaxial strain relaxation occurs and the in-plane lattice constant returns to the original bulk value concomitantly with the introduction of dislocations.[6-8] By studying and assessing $h_c$ in various perovskite oxide thin films and heterostructures, the fundamental correlation between $h_c$ and epitaxial strain would lead to a better understanding of the strain relaxation mechanism in general.

The People-Bean (PB) model is one of the most comprehensive and successful approaches for predicting $h_c$.[9-11] It is a phenomenological model that considers the energies of strain and dislocation within the thin film.[12] It compares the strain energy density, $2G\frac{1+v}{1-v}h\varepsilon^2$ (where $G$ is the shear modulus, $v$ is the Poisson ratio, which is the ratio between the out-of-plane ($\varepsilon_{oop}$) and in-plane lattice mismatch ($\varepsilon_{ip}$), $h$ is the thickness, and $\varepsilon$ is the lattice mismatch; $G$, $v$, and $h$ are the intrinsic values of the thin film), with the dislocation energy density, $\frac{Gb^2}{8\pi\sqrt{2}a_{film}}\ln(\frac{h}{b})$ (where $a_{film}$ is the in-plane lattice constant of the film in the bulk phase, and $b$ is the Burger's vector, which is proportional to $a_{film}$). When the strain energy density exceeds the dislocation energy density at $h_c = \frac{b(1-v)}{40\pi(1+v)}\frac{1}{\varepsilon^2}\ln(\frac{h_c}{b})$, misfit dislocations start to be created with epitaxial strain relaxation. The PB model has been used to successfully predict the $h_c$ of various perovskite oxide thin film systems, including $LaAlO_3$ (LAO) and $PbTiO_3$ thin films on $SrTiO_3$ (STO) substrates and $BaTiO_3$ thin films on Scandate substrates.[13, 14]

The data analytics approach is an emerging tool in materials science and condensed matter physics with practical problem-solving abilities. For example, it can be applied to constructing a magnetic phase diagram by predicting the Néel temperatures of cubic lattices and ferroelectric phase diagram from experimental Raman spectra.[15, 16] The approach was also employed to characterize structural dynamics in glassy liquids and predict the yield strength of high-temperature Cr alloy.[17, 18] Specifically for the case of perovskite oxides, machine learning (ML) was used to predict thermodynamic stabilities,[19] lattice constants,[20] thermal expansion,[21] and the synthesizability of new compounds.[22]

We propose that the approach can be further applied to efficiently identify the correlation between various input features and the $h_c$ of perovskite oxide thin films by considering several different factors that allow going beyond the PB model. Despite the effective predictability of $h_c$, the PB model also has limitations in being universally applied. For example, the PB model often fails to predict $h_c$ in systems with unconventional strain-relaxation characteristics, such as ferroelastic thin films or low mismatched systems.[23, 24] If adequately applied, the data analytics approach will include the concerted effect from various parameters, such as synthesis methods, growth conditions, and type of materials, in determining the epitaxial strain relaxation. Additionally, a quantitative ranking of the relevant parameters in terms of their importance in determining $h_c$ will be possible through correlation analyses. Finally, new augmented functional forms based on combined features can be created to reach high correlations, providing insights in understanding the epitaxial strain relaxation.

In this study, we perform data analytics using correlation analyses and feature engineering to train ML models to understand the epitaxial strain relaxation of perovskite oxide epitaxial thin films. We explain the data analytics process adopted in the current study in Section. 2. We discuss the challenges and limitations in applying the data analytics to actual experimental data, which are inconsistent and sporadic. In Section. 3, we present the data analytics results and discuss the epitaxial strain relaxation mechanism in terms of the PB model. We conclude our study in Section. 4, and briefly explain experimental process for validating the data analytics in Section. 5.

## 2. Data Analytics Process: Challenges and Suggested Resolutions

The data analytics process, illustrated in Figure 1, consists of four steps: (1) Dataset construction, (2) correlation analyses, (3) feature set compilation, and (4) ML model training. Below, we list challenges and resolutions in each step.

### 2.1. Dataset Construction

We collected 82 experimentally reported $h_c$ for the perovskite oxide thin films, as shown in Table S1. Due to the sporadic nature of the data points, we encountered practical challenges in introducing consistent features that comprehensively capture the various experimental conditions. For example, our dataset contains data from six growth methods and 11 different substrates. For the growth of the same $La_{0.7}Sr_{0.3}MnO_3$/LAO (001) system, magnetron sputtering and pulsed laser deposition (PLD) result in drastically different $h_c$ of 2.5 and 12 nm, respectively,[25, 26] highlighting the influence of the growth method on $h_c$. On the other hand, substrates with a significant lattice mismatch or orthorhombic crystal structure would further complicate the analyses. Hence, we compiled a relatively small yet highly consistent dataset by grouping only data points with a similar pedigree (Figure 1a). Our final dataset comprises 23 data points, as shown in Table S2. We have selected the results for the thin films grown on STO, LAO, and $(LaAlO_3)_{0.3}(Sr_2TaAlO_6)_{0.7}$ (LSAT) (001) substrates by PLD.[26-42] Despite small size of the dataset, the following approach was found to be efficient in assessing the epitaxial strain relaxation and predicting $h_c$.

### 2.2. Correlation Analyses

Correlation analyses let us quantitatively examine the contribution of individual features quantitatively and develop physical conclusions (Figure 1b). The analyses identify key physical/chemical features to determine $h_c$ based on two distinct correlation coefficients. Maximal information coefficient (MIC) quantifies nonlinear correlations, and Pearson correlation coefficient (PCC) describes linear correlations with either positive or negative correlations.[43]

### 2.3. Feature Set Compilation

Feature engineering, a process of adjusting features is necessary for achieving realistic and reliable data analytics results.[44, 45] With the information of correlation scores, we adopted various physical hypotheses describing the relation between $h_c$ and epitaxial strain relaxation for the feature set compilation. The optimum feature set found via this process will be used for the ML training. This study classifies features into three categories: ionic, phase, and PB model features (Figure 1b and Table 1). Growth-related parameters, such as thermal expansion coefficient and growth temperature might be considered as important features in determining $h_c$. However, our correlation analyses showed small correlation scores for those parameters, implying that the growth procedure does not strongly affect $h_c$. The application of a physical hypothesis for each feature set (Table 2) is justified as follows. In set A, we speculated the ionic properties, including atomic weight, electron affinity, electronegativity, ionization energy, ionic radius, and oxidation state of individual ions in perovskite structures for the thin film and substrate, might influence $h_c$. Considerations related to the oxidation states of constituent ions were applied to all ionic features. In set B, the general phase features related to epitaxial strain, including $a_{film}$, $v$, and $\varepsilon$, were essential in determining $h_c$. In set C, the PB model features were selected to examine the validity of the PB model, including PB factor, $X_{PB} = a_{film} \frac{1-v}{1+v}\frac{1}{\varepsilon^2}$; strain energy density factor, $E_S = G \frac{1+v}{1-v} \varepsilon^2$; and dislocation energy density factor, $E_D = G\ a_{film}$ ln($a_{film}$). These features were directly adopted from the PB model, but the scale constants were eliminated to reduce them into the simplest numerical form. We also omitted $h$ from the original PB formulas to remove the self-recurring thickness effect. These combined features were expected to provide a concerted approach in understanding $h_c$. Set D includes both PB model features and phase features simultaneously, which lets us examine any synergetic effect between the features.

**2.4. ML Model Training**

ML model training was performed by using an open-source data analytics frontend, Advanced data SCiEnce toolkit for Non-Data Scientist (ASCENDS) (Figure 1d).[46] ML models were trained for a given dataset and feature sets while changing detailed conditions, such as the type of algorithm and scaler, which are intrinsic training parameters. We also tuned the hyperparameter corresponding to the scaler used. Four algorithms, i.e., nearest neighbor (NN) regression, kernel ridge (KR) regression, Bayesian ridge (BR) regression, and support vector machine (SVM), were adopted to train the models.[46] The NN[47], KR[48, 49], BR[50, 51], and SVM[52] regression models were utilized as four representative ML models. NN model employs the results of the k-nearest neighbors' average values for the given data points. The function only takes a portion of the pertinent dataset because it can only be approximated locally. KR is one of the non-parametric forms of ridge regression that combines the kernel technique and ridge regression. It develops a linear model in the implicit feature space caused by the appropriate kernel and data. KR simplifies the computation of inner products in a high-dimensional space by employing the kernel approach. It correlates to a non-linear function in the original space for the non-linear kernels. BR model is a linear-based model, which assumes a relationship between the input and output variables by fitting a linear equation. Instead of employing point estimates, BR formulates a linear relationship using probability distributors. SVM can handle both classification and regression issues. SVM creates a set of hyperplanes in high-dimensional space to classify the data points for a classification task. SVM is more versatile for regression problems by enclosing the function in the ε-insensitive region (ε-tube). To balance model complexity and prediction error in the SVM regression, this tube reformulates the regression problem to identify the function that deviates from the acquired targets throughout all training data the least. ASCENDS saves metadata regarding each training model so that deviation of accuracy ($R^2$) can be calculated by using ten times of trial. The ML model with the highest accuracy was trained using features of high correlation. We have further validated the ML model by comparing its estimated $h_c$ value with the directly obtained experimental $h_c$ value from an example not available in the literature.

## 3. Results and Discussion

The result of correlation analyses (Figure 1b) for the relevant features are shown in Figure 2. Despite the differences between MIC and PCC, $X_{PB}$ and $E_S$ commonly show high correlation scores. The MIC (absolute PCC) scores are 0.687 (0.950) and 0.687 (0.629) for $X_{PB}$ and $E_S$, respectively. In contrast, the correlation scores of $E_D$ are 0.435 and 0.025 for MIC and PCC, respectively, which is significantly lower than those of $X_{PB}$ and $E_S$. This suggests that $X_{PB}$ and $E_S$ are more important than $E_D$, among the PB features. Because $X_{PB}$ and $E_S$ are different from $E_D$ in that they contain both $v$ and $\varepsilon$, it can be further inferred that a combination of $v$ and $\varepsilon$ is critical in constructing the most efficient feature. The result is more intriguing because individual $v$ or $\varepsilon$ alone do not exhibit particularly high correlation scores, yet the combined features of $X_{PB}$ and $E_S$ become the most physically relevant. Note that the overall correlation scores of ionic features in set A (Figure S2) are lower than the ones of set C, and set D with $X_{PB}$, $E_S$.

To compile the features (Figure 1c), we compare the training results of each feature set from feature engineering (Table 2). The results show that the feature sets C and D, including the PB model features, are highly reliable. Figure 3a exhibits the $R^2$ of the ML model for the feature sets presented in Table 2. For set A, all algorithms produced $R^2 < 0.8$. The low $R^2$ values and large deviations imply that the feature set does not represent a valid physical situation. This result is not surprising because the ionic features do not consider any interaction between the film and the substrate. Only the BR algorithm results in a reasonable accuracy for set B, suggesting that set B does not contain enough critical features for predicting $h_c$. Notably, it can be inferred that individual $v$ and $\varepsilon$ are insufficient to construct a valid prediction model. On the other hand, set C exhibits consistently high $R^2$ values, indicating good model training. Set D also shows high $R^2$ values similar to set C. From the results of sets C and D, it is evident that the PB features are crucial in determining $h_c$. Figure 3b presents an example of the ML training results obtained using set C and the BR algorithm, with the highest $R^2$ value of 0.87. The diagonal grey region has a slope of 1, indicating the correspondence of the predicted and actual experimental values of $h_c$. The prediction is encouraging, especially considering experimental uncertainty and sporadic data points, and confirms feature engineering has successfully deduced reliable feature sets.

With the insight from the results of correlation analyses and feature set compilation, we predicted with the trained ML surrogate models (Figure 1d), as summarized in Figure 4. The $h_c$ of the STO/LSAT system is predicted by various ML models based on different feature sets and algorithms. Figure 4a shows the $h_c$ values obtained by the ML models from sets A, B, C, and D (vertical bars) with the BR algorithm. The predictions using feature sets A (42.7 ± 1.5 nm) and B (36.5 ± 6.2 nm) largely underestimates $h_c$. On the other hand, using feature sets C (78.8 ± 1.0 nm) and D (77.2 ± 4.2 nm) the predictions lie just beneath the PB model calculation result (86.4 nm, red horizontal dashed line). We further compare the algorithm-dependent results of using sets C (Figure 4b) and D (Figure 4c). Set C produces more consistent results with less variation among different algorithms. This reiterates that the individual features of $v$ and $\varepsilon$ included in set D might obscure the effective model construction. Their augmented form is essential in understanding the epitaxial strain relaxation.

For more realistic validation of our ML model, we compared our results to the experimental result. The X-ray reflectivity (XRR) results of four samples with thicknesses of 37.2, 72.0, 88.5, and 117.0 nm are shown in Figure S1. X-ray diffraction reciprocal space map (XRD-RSM) measurements were taken for the samples to investigate the epitaxial strain relaxation. As the epitaxial strain relaxes with increasing thickness, additional Bragg peaks (i.e., relaxed regions) emerge, breaking the mirror symmetry of the original Bragg peak (i.e., strained region). The upper panels in Figure S1c show the RSMs around the substrate LSAT (103) and the film STO (103) peaks. The regions marked by white boxes are magnified in the lower panels to assess the strain relaxation of the STO thin film in further detail. The peaks are symmetric for the 37.2 and 72.0 nm films, but they become progressively asymmetric for the 88.5 and 117.0 nm films, suggesting that the strain relaxation occurs between 72.0 and 88.5 nm. This strain relaxation behavior was further quantitatively examined using a bi-Gaussian fitting of the STO (103) peak (Figure 5 and S2). Bi-Gaussian function, which has a distinct standard deviation for the left ($W_1$) and right half ($W_2$) of the peak, is an effective tool for quantifying the asymmetry that originates from strain relaxation. The line profiles through the black lines in the lower panels of Figure S1c and their bi-Gaussian fittings (red lines) are plotted in Figure S3a. As shown in Figure 5, $(W_1 - W_2)/W_2$, the normalized difference between the width on the left and right side increases dramatically as the film thickness ≥ 88.5 nm, the thickness near which the strain relaxation begins. Therefore, the $h_c$ of the STO/LSAT system was experimentally determined to be 72.0 – 88.5 nm and it is consistent with our ML model result.

Whereas the original PB model show decent prediction of $h_c$ as expected (Fig. S4a), the data analytics approach provides hidden insight of the epitaxial strain relaxation of perovskite oxide thin film system. Particularly, we note that the strong correlations between $h_c$ and various features are not evident when $h_c$ values from the literature are directly plotted (Figure S4). For example, Figures S4b-d show $h_c$ values plotted as functions of $E_S$, $v$, and $\varepsilon$. This emphasizes the merit of applying data analytics which quantitatively characterize the augmented features of $X_{PB}$ and $E_S$ to be essential in understanding the epitaxial strain relaxation. As discussed previously, both $X_{PB}$ and $E_S$ include the parameters $v$ and $\varepsilon$, yet feature sets including pure $v$ and $\varepsilon$, do not result in particularly high precision in predicting $h_c$. Physically, this might imply that independent information on either the value of the in-plane lattice structure, $\varepsilon$, or the relationship between the in-plane and out-of-plane lattice structure, $v$, does not provide meaningful understanding of the epitaxial strain relaxation. Again, the augmented features of $X_{PB}$ and $E_S$ are critical, as the elastic modulation of the thin film should be interpreted as a three-dimensional phenomenon.

## 4. Conclusion

We demonstrate the feasibility of establishing a streamlined data analytics workflow to efficiently evaluate and introduce relevant features that capture the physics of strain relaxation in epitaxial thin films. First, we collected various experimental $h_c$ data which are sporadic in nature. Second, we augmented physical/chemical features for detailed correlation analyses. Third, we refined the dataset into consistent sub-dataset by applying the result of correlation analyses and prevailing physical conditions, which inevitably reduced our dataset. Despite the small number of sporadic data, our carefully chosen conditions were proven to be highly consistent. Consequently, the data analytics process presented in the current study based on the PB model provides an obvious first step for understanding $h_c$ by quantitatively identifying key features (i.e., the Poisson's ratio $\nu$ and the lattice mismatch $\varepsilon$) that affect the epitaxial strain relaxation. We experimentally validated the predicted $h_c$ of STO thin films grown on LSAT (001) substrates, showing a good agreement. This study introduces challenges in ML approach using sporadic experimental dataset and proposes its systematic solutions. By doing so, we emphasize that using refined dataset within the context of modern data analytics can help achieving a better understanding. In particular, the quantitative analyses of ML successfully provide us with the unique physical insight about three-dimensional nature of epitaxial strain relaxation mechanism intuitively by focusing on the key features. Initiating a framework for understanding the epitaxial strain relaxation would inspire the community to consistently collect/compile the dataset. Furthermore, we anticipate that the demonstrated data analytics approach can be further applied beyond the example used in the present study.

## 5. Experimental Section

We experimentally fabricated epitaxial STO thin films on LSAT substrates and determined the actual range of $h_c$ to validate the data analytics approach (Figure S1). The system was selected because it was not available in the literature, so pure prediction is possible. STO thin films were fabricated on LSAT (001) substrate using PLD. We grew the film at 750 °C and 100 mTorr of $O_2$ partial pressure, using a KrF excimer laser (248 nm; IPEX-868,

Lightmachinery) with 1.5 J $cm^{-2}$ of fluence and 5 Hz of repetition rate. The thicknesses of the STO thin films were determined by XRR (PANalytical X'Pert and a Rigaku Smartlab XRD), as the films had atomically sharp surfaces and interfaces.

**Acknowledgements**

This work was supported by the Basic Science Research Programs through the National Research Foundation of Korea (NRF) (NRF-2021R1A2C201134012).

**ORCID**

*Woo Seok Choi* https://orcid.org/0000-0002-2872-6191

**References**

1. B. Kim, P. Liu, J. M. Tomczak, and C. Franchini. Strain-Induced Tuning of the Electronic Coulomb Interaction in *3d* Transition Metal Oxide Perovskites. Phys. Rev. B. 2018; 98(7):075130.
2. J. Cao and J. Wu. Strain Effects in Low-Dimensional Transition Metal Oxides. Mater. Sci. Eng. R Rep. 2011; 71(2):35-52.
3. D. Pesquera, G. Herranz, A. Barla, E. Pellegrin, F. Bondino, E. Magnano, F. Sánchez, and J. Fontcuberta. Surface Symmetry-Breaking and Strain Effects on Orbital Occupancy in Transition Metal Perovskite Epitaxial Films. Nat. Commun. 2012; 3(1):1189.
4. D. J. Dunstan. Strain and Strain Relaxation in Semiconductors. J. Mater. Sci.: Mater. Electron. 1997; 8(6):337-75.

5. Y. Chen, Y. Lei, Y. Li, Y. Yu, J. Cai, M.-H. Chiu, R. Rao, Y. Gu, C. Wang, W. Choi, H. Hu, C. Wang, Y. Li, J. Song, J. Zhang, B. Qi, M. Lin, Z. Zhang, A. E. Islam, B. Maruyama, S. Dayeh, L.-J. Li, K. Yang, Y.-H. Lo, and S. Xu. Strain Engineering and Epitaxial Stabilization of Halide Perovskites. Nature. 2020; 577(7789):209-15.

6. D. G. Schlom, L.-Q. Chen, C.-B. Eom, K. M. Rabe, S. K. Streiffer, and J.-M. Triscone. Strain Tuning of Ferroelectric Thin Films. Annu. Rev. Mater. Res. 2007; 37(1):589-626.

7. A. E. Romanov and J. S. Speck. Stress Relaxation in Mismatched Layers due to Threading Dislocation Inclination. Appl. Phys. Lett. 2003; 83(13):2569-71.

8. D. Holec, Y. Zhang, D. V. S. Rao, M. J. Kappers, C. McAleese, and C. J. Humphreys. Equilibrium Critical Thickness for Misfit Dislocations in III-Nitrides. J. Appl. Phys. 2008; 104(12):123514.

9. J. W. Matthews and A. E. Blakeslee. Defects in Epitaxial Multilayers: I. Misfit Dislocations. J. Cryst. Growth. 1974; 27:118-25.

10. A. Fischer, H. Kühne, and H. Richter. New Approach in Equilibrium Theory for Strained Layer Relaxation. Phys. Rev. Lett. 1994; 73(20):2712-15.

11. L. B. Freund. The Stability of a Dislocation Threading a Strained Layer on a Substrate. J. Appl. Mech. 1987; 54(3):553-57.

12. R. People and J. C. Bean. Calculation of Critical Layer Thickness Versus Lattice Mismatch for $Ge_xSi_{1-x}$/Si Strained-Layer Heterostructures. Appl. Phys. Lett. 1985; 47(3):322-24.

13. S. Venkatesan, A. Vlooswijk, B. J. Kooi, A. Morelli, G. Palasantzas, J. T. M. De Hosson, and B. Noheda. Monodomain Strained Ferroelectric $PbTiO_3$ Thin Films: Phase Transition and Critical Thickness Study. Phys. Rev. B. 2008; 78(10).

14. Y. B. Chen, H. P. Sun, M. B. Katz, X. Q. Pan, K. J. Choi, H. W. Jang, and C. B. Eom. Interface Structure and Strain Relaxation in $BaTiO_3$ Thin Films Grown on $GdScO_3$ and $DyScO_3$ Substrates with Buried Coherent $SrRuO_3$ Layer. Appl. Phys. Lett. 2007; 91(25).

15. K. Ch'ng, J. Carrasquilla, R. G. Melko, and E. Khatami. Machine Learning Phases of Strongly Correlated Fermions. Phys. Rev. X. 2017; 7(3):031038.

16. A. Cui, K. Jiang, M. Jiang, L. Shang, L. Zhu, Z. Hu, G. Xu, and J. Chu. Decoding Phases of Matter by Machine-Learning Raman Spectroscopy. Phys. Rev. Appl. 2019; 12(5):054049.

17. S. S. Schoenholz, E. D. Cubuk, D. M. Sussman, E. Kaxiras, and A. J. Liu. A structural approach to relaxation in glassy liquids. Nat. Phys. 2016; 12(5):469-71.

18. J. Peng, Y. Yamamoto, J. A. Hawk, E. Lara-Curzio, and D. Shin. Coupling Physics in Machine Learning to Predict Properties of High-Temperatures Alloys. Npj Comput. Mater. 2020; 6(1):141.

19. W. Li, R. Jacobs, and D. Morgan. Predicting the Thermodynamic Stability of Perovskite Oxides Using Machine Learning Models. Comp. Mater. Sci. 2018; 150:454-63.

20. Y. Zhang and X. Xu. Machine Learning Lattice Constants for Cubic Perovskite $A_2^{2+}BB'O_6$ Compounds. CrystEngComm. 2020; 22(38):6385-97.

21. J. Peng, N. S. Harsha Gunda, C. A. Bridges, S. Lee, J. Allen Haynes, and D. Shin. A Machine Learning Approach to Predict Thermal Expansion of Complex Oxides. Comp. Mater. Sci. 2021; 210:111034.

22. P. V. Balachandran, A. A. Emery, J. E. Gubernatis, T. Lookman, C. Wolverton, and A. Zunger. Predictions of new ABO(3) perovskite compounds by combining machine learning and density functional theory. Phys Rev Mater. 2018; 2(4).

23. S. Yoon, X. Gao, J. M. Ok, Z. Liao, M.-G. Han, Y. Zhu, P. Ganesh, M. F. Chisholm, W. S. Choi, and H. N. Lee. Strain-Induced Atomic-Scale Building Blocks for Ferromagnetism in Epitaxial LaCoO3. Nano Lett. 2021; 21(9):4006-12.

24. T. Kujofsa and J. E. Ayers. Critical Layer Thickness: Theory and Experiment in the ZnSe/GaAs (001) Material System. International Journal of High Speed Electronics and Systems. 2017; 26(03):1740020.

25. J. Santiso, J. Roqueta, N. Bagués, C. Frontera, Z. Konstantinovic, Q. Lu, B. Yildiz, B. Martínez, A. Pomar, L. Balcells, and F. Sandiumenge. Self-Arranged Misfit Dislocation Network Formation upon Strain Release in $La_{0.7}Sr_{0.3}MnO_3/LaAlO_3(100)$ Epitaxial Films under Compressive Strain. ACS Appl. Mater. Interfaces. 2016; 8(26):16823-32.

26. A. Tebano, C. Aruta, P. G. Medaglia, F. Tozzi, G. Balestrino, A. A. Sidorenko, G. Allodi, R. De Renzi, G. Ghiringhelli, C. Dallera, L. Braicovich, and N. B. Brookes. Strain-Induced Phase Separation in $La_{0.7}Sr_{0.3}MnO_3$ Thin Films. Phys. Rev. B. 2006; 74(24):245116.

27. H. L. Ju, K. M. Krishnan, and D. Lederman. Evolution of Strain-Dependent Transport Properties in Ultrathin $La_{0.67}Sr_{0.33}MnO_3$ Films. J. Appl. Phys. 1998; 83(11):7073-75.

28. S. H. Lim, M. Murakami, W. L. Sarney, S. Q. Ren, A. Varatharajan, V. Nagarajan, S. Fujino, M. Wuttig, I. Takeuchi, and L. G. Salamanca-Riba. The Effects of Multiphase Formation on Strain Relaxation and Magnetization in Multiferroic $BiFeO_3$ Thin Films. Adv. Funct. Mater. 2007; 17(14):2594-99.

29. J. L. Maurice, F. Pailloux, A. Barthélémy, O. Durand, D. Imhoff, R. Lyonnet, A. Rocher, and J. P. Contour. Strain Relaxation in the Epitaxy of $La_{2/3}Sr_{1/3}MnO_3$ Grown by Pulsed-Laser Deposition on $SrTiO_3(001)$. Philos. Mag. Lett. 2003; 83(28):3201-24.

30. Y. H. Chu, T. Zhao, M. P. Cruz, Q. Zhan, P. L. Yang, L. W. Martin, M. Huijben, C. H. Yang, F. Zavaliche, H. Zheng, and R. Ramesh. Ferroelectric Size Effects in Multiferroic $BiFeO_3$ Thin Films. Appl. Phys. Lett. 2007; 90(25):252906.

31. P. Mirzadeh Vaghefi, A. Baghizadeh, M. G. Willinger, M. J. Pereira, D. A. Mota, B. G. Almeida, J. Agostinho Moreira, and V. S. Amaral. Thickness Dependence of Microstructure in Thin $La_{0.7}Sr_{0.3}MnO_3$ Films Grown on (1 0 0) $SrTiO_3$ Substrate. J. Phys. D Appl. Phys. 2017; 50(39):395301.

32. L. Ranno, A. Llobet, R. Tiron, and E. Favre-Nicolin. Strain-Induced Magnetic Anisotropy in Epitaxial Manganite Films. Appl. Surf. Sci. 2002; 188(1):170-75.

33. L. Qiao, T. C. Droubay, T. Varga, M. E. Bowden, V. Shutthanandan, Z. Zhu, T. C. Kaspar, and S. A. Chambers. Epitaxial Growth, Structure, and Intermixing at the $LaAlO_3/SrTiO_3$ Interface as The Film Stoichiometry is Varied. Phys. Rev. B. 2011; 83(8):085408.

34. T. Suzuki, Y. Nishi, and M. Fujimoto. Analysis of Misfit Relaxation in Heteroepitaxial $BaTiO_3$ Thin Films. Philos. Mag. Lett. A. 1999; 79(10):2461-83.

35. A. Visinoiu, M. Alexe, H. N. Lee, D. N. Zakharov, A. Pignolet, D. Hesse, and U. Gösele. Initial Growth Stages of Epitaxial $BaTiO_3$ Films on Vicinal $SrTiO_3$ (001) Substrate Surfaces. J. Appl. Phys. 2002; 91(12):10157-62.

36. R. Guo, L. Shen, H. Wang, Z. Lim, W. Lu, P. Yang, Ariando, A. Gruverman, T. Venkatesan, Y. P. Feng, and J. Chen. Tailoring Self-Polarization of $BaTiO_3$ Thin Films by Interface Engineering and Flexoelectric Effect. Adv. Mater. Interfaces. 2016; 3(23):1600737.

37. J. Zhu, X. H. Wei, Y. Zhang, and Y. R. Li. Study on Interfacial Strain Behavior of Functional Oxide Heterostructures. J. Appl. Phys. 2006; 100(10):104106.

38. M. Fujimoto. Defects in Epitaxially Grown Perovskite Thin Films. J. Cryst. Growth. 2002; 237-239:430-37.

39. G. H. Lee, B. C. Shin, and I. S. Kim. Critical Thickness of $BaTiO_3$ Film on $SrTiO_3$ (001) Evaluated by Reflection High-Energy Electron Diffraction. Mater. Lett. 2001; 50(2):134-37.

40. H. J. Lee, S. S. Lee, J. H. Kwak, Y.-M. Kim, H. Y. Jeong, A. Y. Borisevich, S. Y. Lee, D. Y. Noh, O. Kwon, Y. Kim, and J. Y. Jo. Depth Resolved Lattice-Charge Coupling in Epitaxial $BiFeO_3$ Thin Film. Sci. Rep. 2016; 6(1):38724.

41. Y. S. Kim, D. H. Kim, J. D. Kim, Y. J. Chang, T. W. Noh, J. H. Kong, K. Char, Y. D. Park, S. D. Bu, J.-G. Yoon, and J.-S. Chung. Critical Thickness of Ultrathin Ferroelectric $BaTiO_3$ Films. Appl. Phys. Lett. 2005; 86(10):102907.

42. D. Zhang, Y. Wang, N. Lu, X. Sui, Y. Xu, P. Yu, and Q.-K. Xue. Origin of the Anomalous Hall Effect in $SrCoO_3$ Thin Films. Phys. Rev. B. 2019; 100(6):060403.

43. D. N. Reshef, Y. A. Reshef, H. K. Finucane, S. R. Grossman, G. McVean, P. J. Turnbaugh, E. S. Lander, M. Mitzenmacher, and P. C. Sabeti. Detecting Novel Associations in Large Data Sets. Science. 2011; 334(6062):1518.

44. P. Domingos. A few Useful Things to Know about Machine Learning. Commun. ACM. 2012; 55(10):78-87.

45. D. Dai, T. Xu, X. Wei, G. Ding, Y. Xu, J. Zhang, and H. Zhang. Using Machine Learning and Feature Engineering to Characterize Limited Material Datasets of High-Entropy Alloys. Comp. Mater. Sci. 2020; 175:109618.

46. J. Peng, S. Lee, A. Williams, J. A. Haynes, and D. Shin. Advanced Data Science Toolkit for Non-Data Scientists – A User Guide. Calphad. 2020; 68:101733.

47. N. S. Altman. An Introduction to Kernel and Nearest-Neighbor Nonparametric Regression. Amer. Statist. 1992; 46(3):175-85.

48. A. E. Hoerl and R. W. Kennard. Ridge Regression: Applications to Nonorthogonal Problems. Technometrics. 1970; 12(1):69-82.

49. A. Tikhonov, A. Goncharsky, V. Stepanov, and A. Yagola. Numerical methods for the solution of ill-posed problems: Springer Science & Business Media. 2013.

50. D. J. C. Mackay. Bayesian Interpolation. Neural Comput. 1992; 4:415-47.

51. M. E. Tipping. Sparse bayesian learning and the relevance vector machine. J. Mach. Learn. Res. 2001; 1:211–44.

52. R. Khanna and M. Awad, "Efficient Learning Machines: Theories, Concepts, and Applications for Engineers and System Designers.″Apress, (2015).

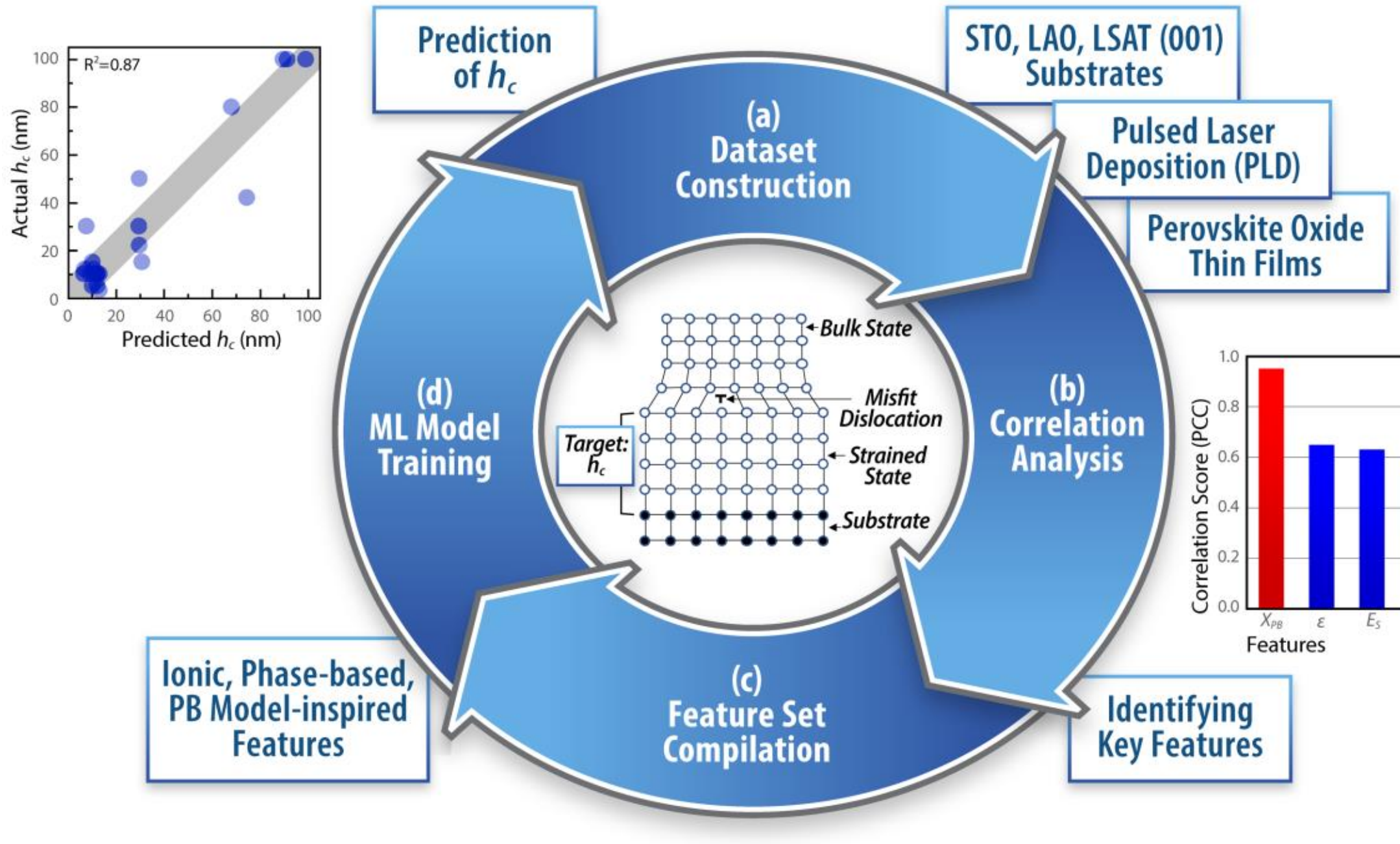

**FIGURE 1.** Schematic workflow of modern data analytics for predicting $h_c$. a) Experimental $h_c$ values of perovskite oxide thin films on STO, LAO, and LSAT (001) substrates fabricated by PLD are collected for the dataset construction. b) Quantitative correlation analyses provide insight into epitaxial strain relaxation by highlighting the underlying correlation. c) Various physical features and feature sets were examined and constructed to create an ideal feature for data analytics. d) Model prediction was applied to predict $h_c$ and compare it with the experimental value.

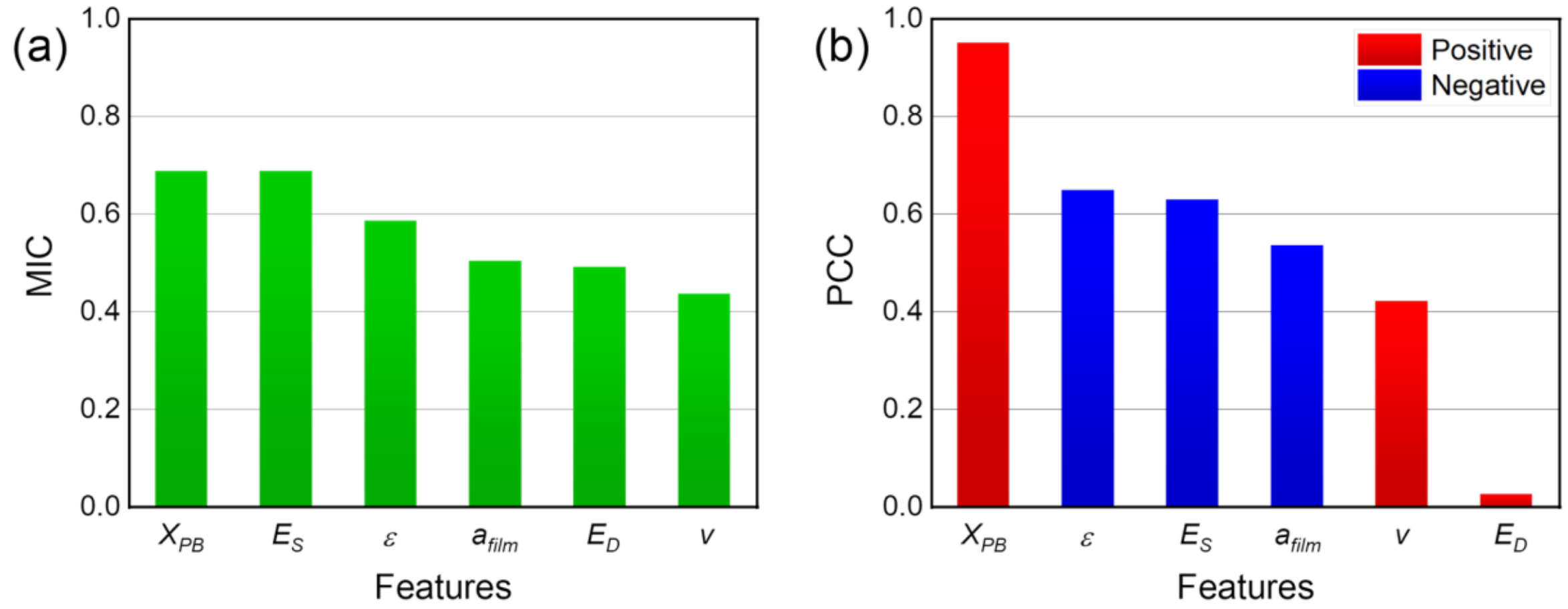

**FIGURE 2.** Correlation analyses in MIC and PCC for the features in set D. Correlation coefficients in a) MIC and b) PCC. For the PCC, red (blue) bars represent positive (negative) correlations.

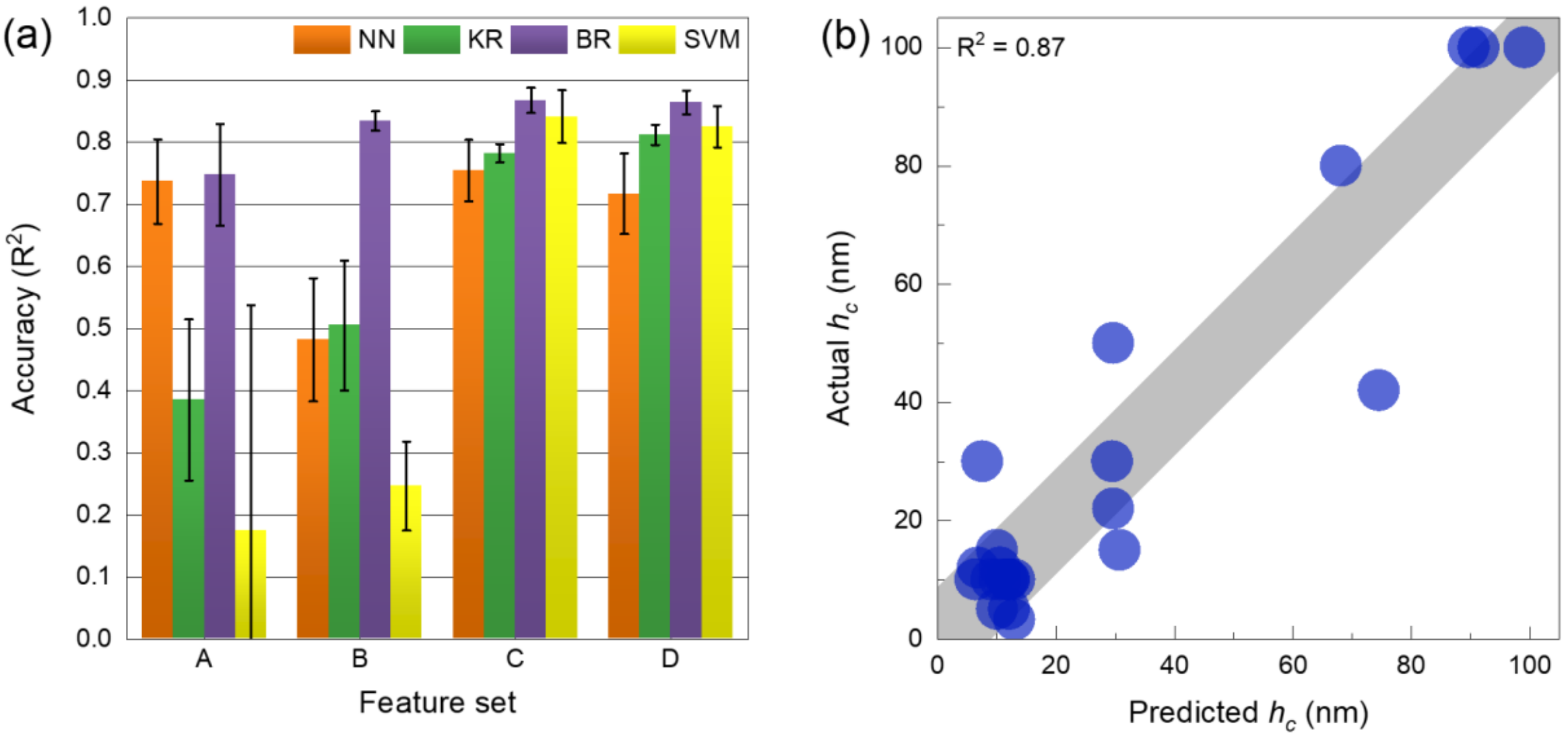

**FIGURE 3.** Accuracy ($R^2$) of each feature set sorted by different algorithms. a) The graph shows the $R^2$ values of the model training using four algorithms, NN, KR, BR, and SVM. b) The ML training result of set C using the BR algorithm with the highest $R^2$ value. Each point represents experimental data with $h_c$ values with their predicted values. The thick gray line represents a slope of 1. BR, Bayesian ridge; KR, kernel ridge; NN, nearest neighbor; SVM, support vector machine.

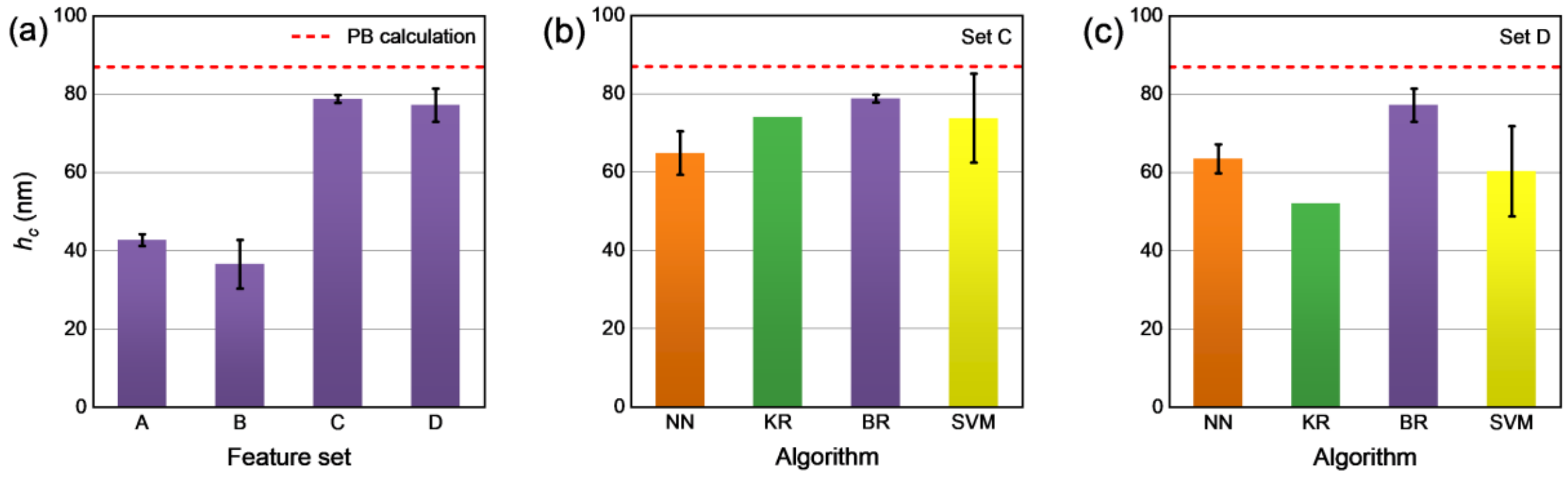

**FIGURE 4.** Comparison between the ML estimated, PB model calculated, and experimental values of $h_c$ for the STO/LSAT system. a) $h_c$ values with error bars (standard deviation) obtained by the ML models using sets A, B, C, and D (purple vertical bars) with the best-performing BR algorithm. $h_c$ values with error bars (standard deviation) obtained by the ML models using b) set C and c) set D based on different algorithms (colored bars). Direct PB model calculation (red horizontal dashed line) are also shown for comparison.

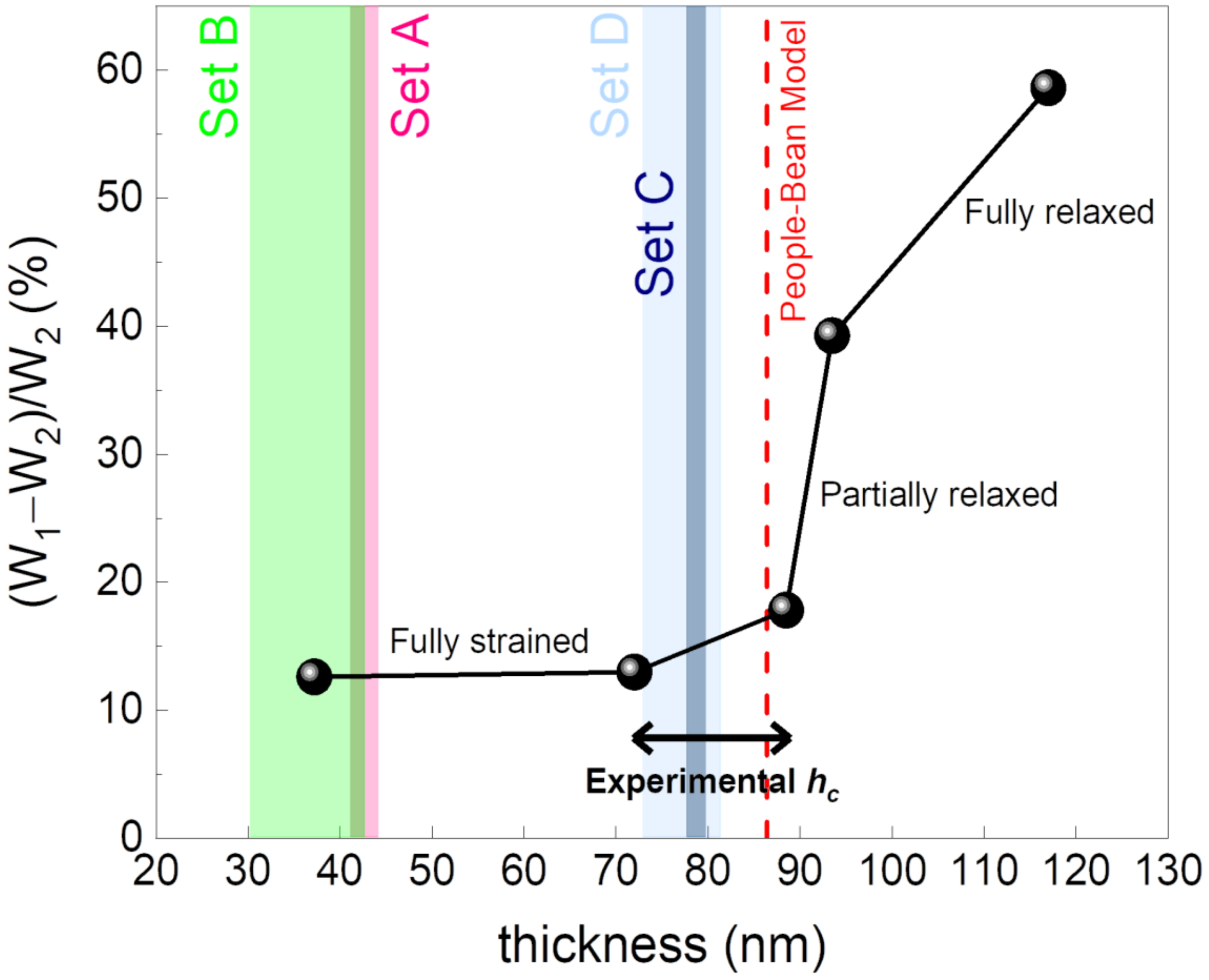

**FIGURE 5.** Asymmetrical line profiles of the STO film peak and epitaxial strain relaxation. Black solid circles represent experimental $(W_1 - W_2)/W_2$ values plotted as a function of the film thickness, quantitatively characterizing the peak asymmetry, and hence, the epitaxial strain

relaxation. The vertical bars represent the ML predicted $h_c$ values from various feature sets, where their thicknesses correspond to the error bars. The $h_c$ values of Set C and Set D lie well between experimental $h_c$ region indicated by an arrow.

| **Ionic features** | | |
|---|---|---|
| Atomic weight | | |
| Electron affinity | | |
| Electronegativity | | |
| Ionization energy | | |
| Ionic radius | | |
| Oxidation states | | |
| **Phase features** | **Definition** | **Symbol** |
| Lattice constant of film | $a_{film}$ | $a_{film}$ |
| Lattice mismatch | $\frac{a_{film} - a_{sub}}{a_{sub}}$ | $\varepsilon$ |
| Poisson ratio | $-\frac{\varepsilon_{oop}}{\varepsilon_{ip}}$ | $v$ |
| **PB model features** | **Definition** | **Symbol** |
| PB factor | $a_{film} \frac{1-v}{1+v} \frac{1}{\varepsilon^2}$ | $X_{PB}$ |
| Strain energy density factor | $G \frac{1+v}{1-v} \varepsilon^2$ | $E_S$ |
| Dislocation energy density factor | $G\, a_{film} \ln(a_{film})$ | $E_D$ |

**TABLE 1**. Description of features used in data analytics. Definitions and symbols of each individual and combined feature used for data analytics. The features were classified into three types: ionic features, phase features, and PB model features.

| Feature set | Features |
|---|---|
| A | Ionic features |
| B | $a_{film}$, $v$, $\varepsilon$ |
| C | $X_{PB}$, $E_s$, $E_D$ |
| D | $X_{PB}$, $E_s$, $E_D$, $a_{film}$, $v$, $\varepsilon$ |

**TABLE 2.** Corresponding features of each feature set. The table lists the four feature sets using features that correspond under different physical hypotheses: ionic features set A, phase features set B, PB features set C, combined features set D.

## SUPPORTING INFORMATION

Additional supporting information may be found in the online version of the article at the publisher's website.

**Supplemental Materials**

# Application of Machine Learning to Sporadic Experimental Data for Understanding Epitaxial Strain Relaxation

*Jin Young Oh*[1], *Dongwon Shin*[2,*], *and Woo Seok Choi*[1,*]

[1]*Department of Physics, Sungkyunkwan University, Suwon 16419, Republic of Korea*
[2]*Materials Science and Technology Division, Oak Ridge National Laboratory, Oak Ridge, TN 37831, USA*

Corresponding author e-mail : shind@ornl.gov, choiws@skku.edu

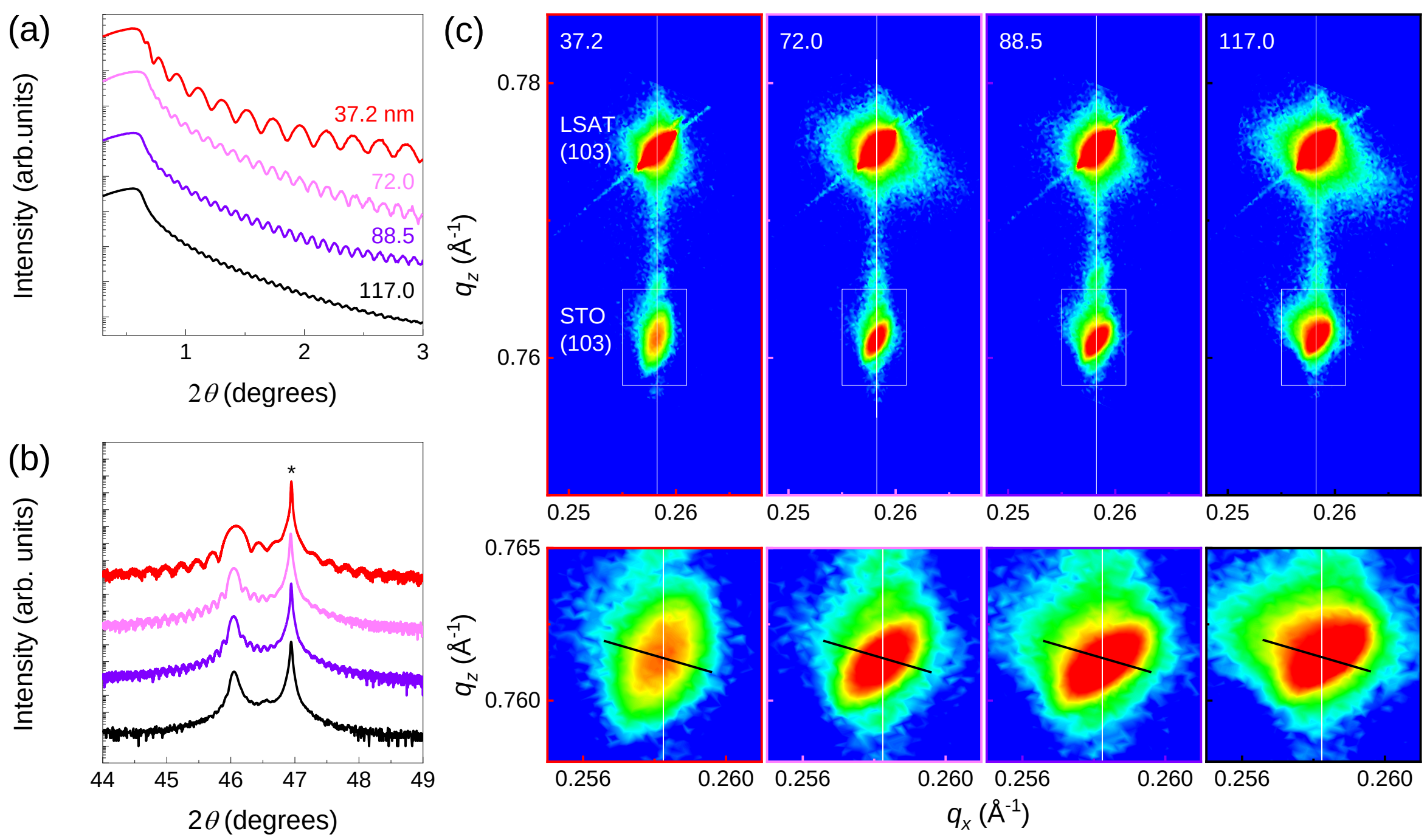

**FIGURE S1.** Characterization of the strain state of STO thin films on LSAT substrate. (a) XRR, (b) $\theta$-$2\theta$, and (c) XRD-RSM data for epitaxial STO thin films on LSAT (001) substrates with representative thicknesses. The asterisk in (b) indicates the LSAT (002) substrate peak and the peak around 46.1° marks the STO (002) peak.

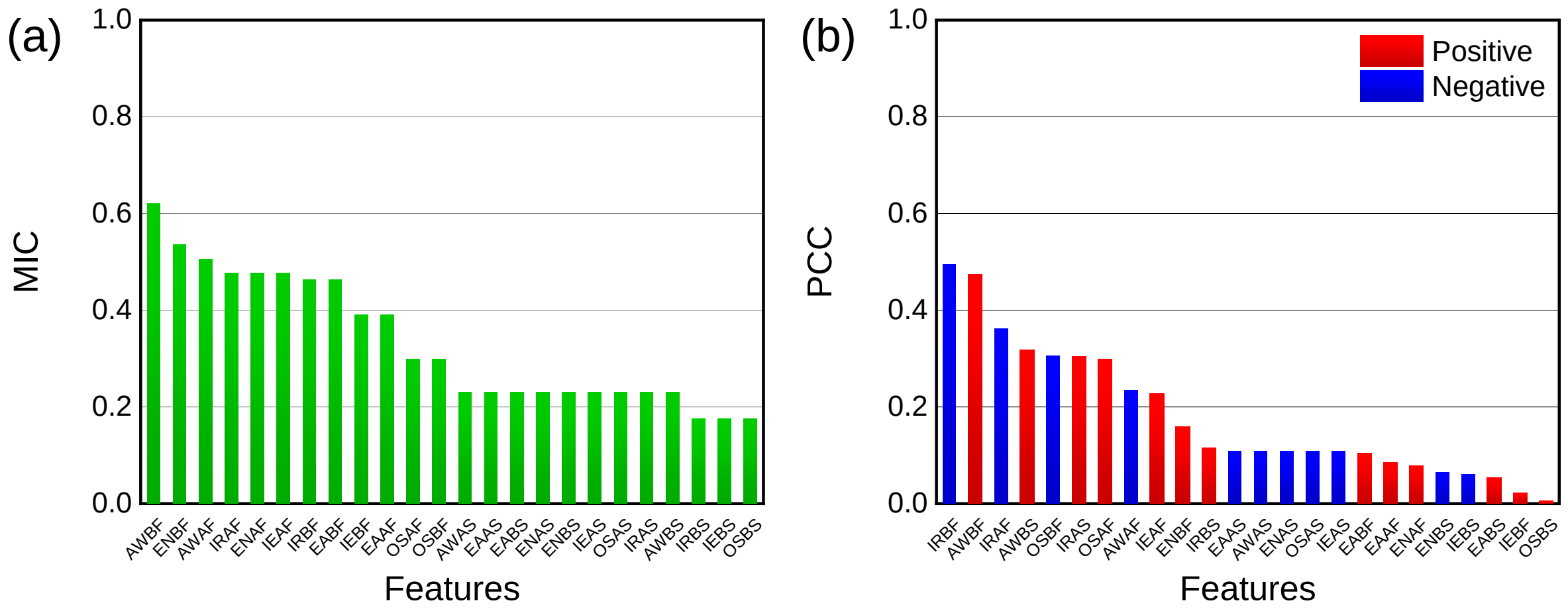

**FIGURE S2.** Correlation analyses in MIC and PCC for the features in set A. Correlation coefficients for set A, ionic feature set. Overall correlation scores are relatively lower than set C, and Set D.

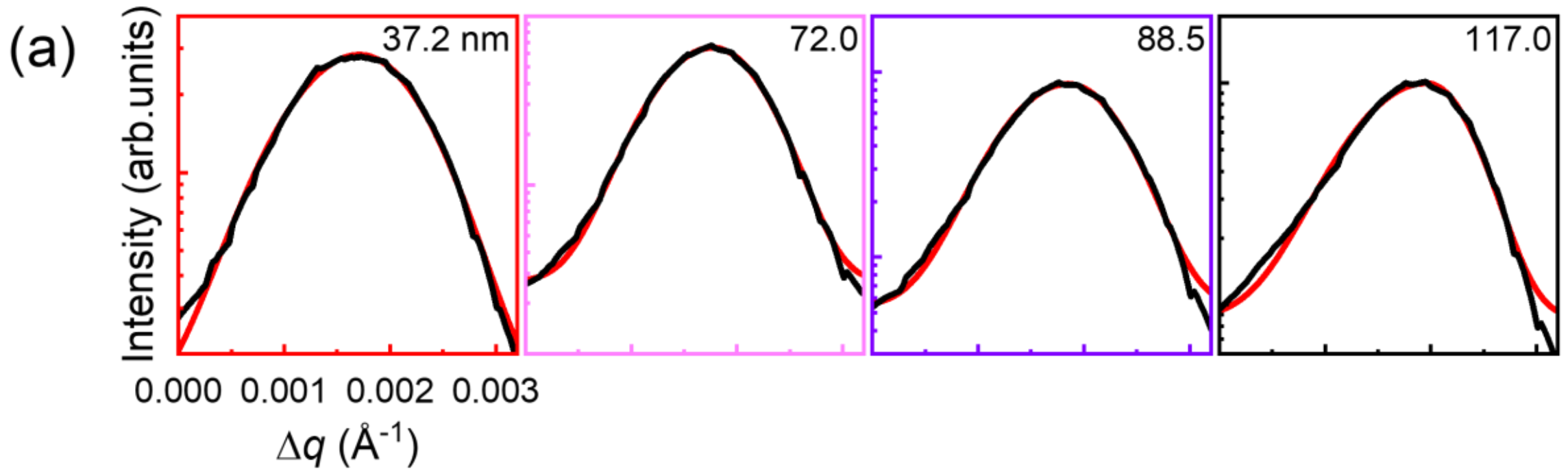

(b)

| Thickness (nm) | 37.2 | 72.0 | 88.5 | 117.0 |
| --- | --- | --- | --- | --- |
| $W_1$ ($10^{-4} \cdot Å^{-1}$) | 6.61 | 4.87 | 5.42 | 6.33 |
| $W_2$ ($10^{-4} \cdot Å^{-1}$) | 5.87 | 4.31 | 4.60 | 3.99 |
| $(W_1 - W_2)/W_2$ | **12.61%** | **12.99%** | **17.82%** | **58.64%** |

**FIGURE S3.** Detailed information of the asymmetric plot. (a) Line profiles crossing the STO (103) peak and bi-Gaussian fitting. (b) The table summarizes the $(W_1 - W_2)/W_2$ values.

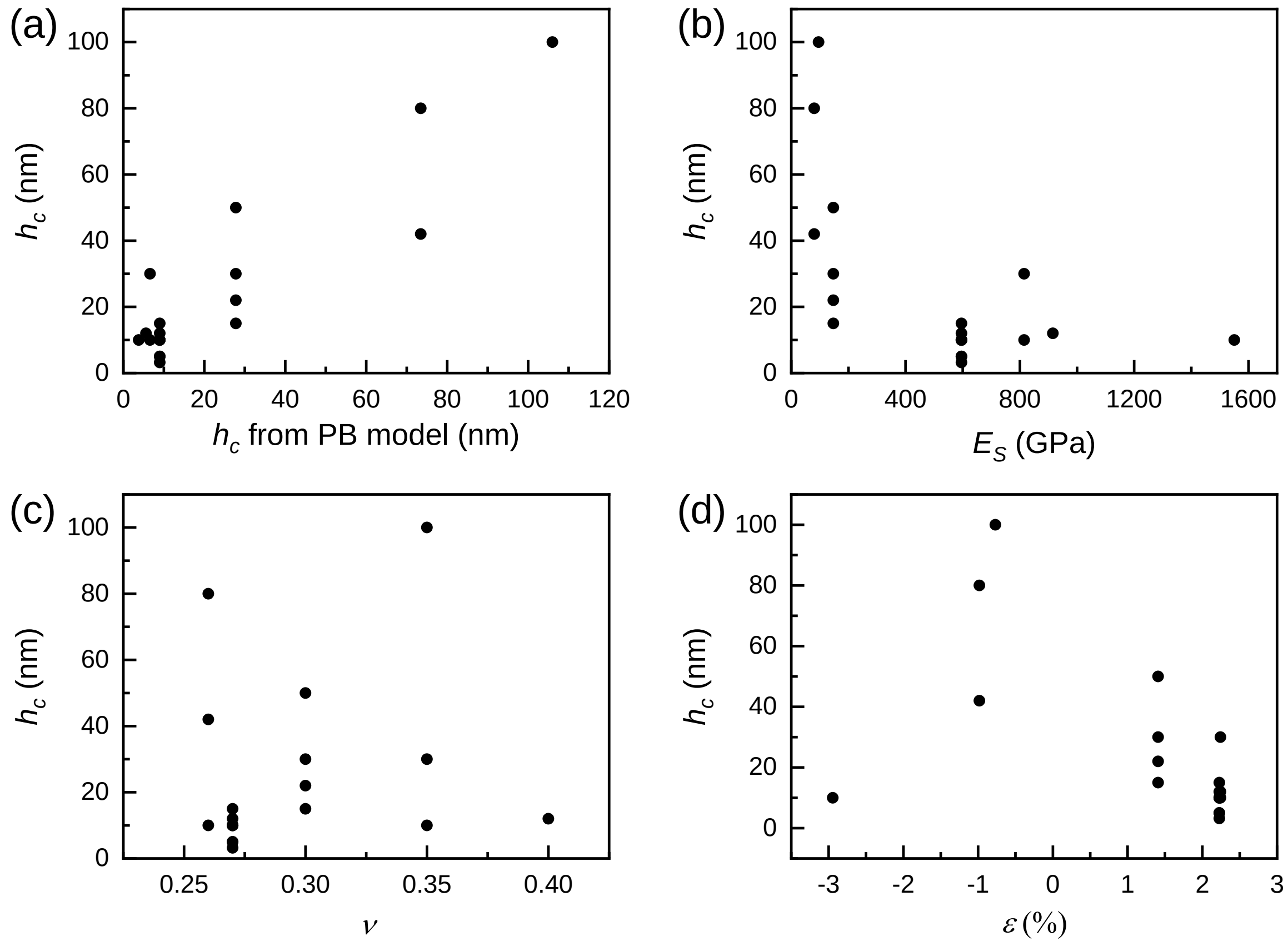

**FIGURE S4.** Correlation between $h_c$ and various features. A plot of actual $h_c$ with respect to various features, including (a) $h_c$ from PB model, (b) $E_s$ (c) $\nu$, and (d) $\varepsilon$.

| Author/year | DOI link | Thin film | Substrate | $h_c$ |
|---|---|---|---|---|
| H.L.Ju 1998 | 10.1063/1.367864 | $La_{0.66}Sr_{0.33}MnO_3$ | $LaAlO_3$ | 10 |
| H.L.Ju 1998 | 10.1063/1.367864 | $La_{0.66}Sr_{0.33}MnO_3$ | $LaAlO_3$ | 30 |
| A.Tebano 2006 | 10.1103/PhysRevB.74.245116 | $La_{0.7}Sr_{0.3}MnO_3$ | $LaAlO_3$ | 12 |
| S.H.Lim 2007 | 10.1002/adfm.200700055 | $BiFeO_3$ | $SrTiO_3$ | 50 |
| Y.H.Chu 2007 | 10.1063/1.2750524 | $BiFeO_3$ | $SrTiO_3$ | 30 |
| S.Geprags 2011 | http://mediatum.ub.tum.de/?id=1091602 | $BiFeO_3$ | $SrTiO_3$ | 22 |
| J.L.Maurice 2003 | 10.1080/14786430310001603436 | $La_{0.66}Sr_{0.33}MnO_3$ | $SrTiO_3$ | 100 |
| P.M.Vaghefi 2017 | 10.1088/1361-6463/aa80bf | $La_{0.7}Sr_{0.3}MnO_3$ | $SrTiO_3$ | 100 |
| L. Ranno 2002 | 10.1016/S0169-4332(01)00730-9 | $La_{0.7}Sr_{0.3}MnO_3$ | $SrTiO_3$ | 100 |
| L.Qiao 2011 | 10.1103/PhysRevB.83.085408 | $LaAlO_3$ | $SrTiO_3$ | 10 |
| T.Suzuki 1999 | 10.1080/01418619908214294 | $BaTiO_3$ | $SrTiO_3$ | 12 |
| A.Visinoiu 2002 | 10.1063/1.1478800 | $BaTiO_3$ | $SrTiO_3$ | 5 |
| R.Guo 2016 | 10.1002/admi.201600737 | $BaTiO_3$ | $SrTiO_3$ | 10 |
| J.Zhu 2006 | 10.1063/1.2375014 | $BaTiO_3$ | $SrTiO_3$ | 3.2 |
| M.Fujimoto 2002 | 10.1016/S0022-0248(01)01962-5 | $BaTiO_3$ | $SrTiO_3$ | 5 |
| M.Fujimoto 2002 | 10.1016/S0022-0248(01)01962-5 | $BaTiO_3$ | $SrTiO_3$ | 10 |
| G.H.Lee 2001 | 10.1016/S0167-577X(00)00430-4 | $BaTiO_3$ | $SrTiO_3$ | 10 |
| H.I.Seo 2020 | To be published | $BaTiO_3$ | $SrTiO_3$ | 10 |
| H.I.Seo 2020 | To be published | $BaTiO_3$ | $SrTiO_3$ | 15 |
| H.J.Lee 2016 | 10.1038/srep38724 | $BiFeO_3$ | $SrTiO_3$ | 15 |
| Y.S.Kim 2005 | 10.1063/1.1880443 | $BaTiO_3$ | $SrTiO_3$ | 10 |
| D.Zhang 2019 | 10.1103/PhysRevB.100.060403 | $SrCoO_3$ | LSAT | 42 |
| D.Zhang 2019 | 10.1103/PhysRevB.100.060403 | $SrCoO_3$ | LSAT | 80 |
| L.Ranno 2002 | 10.1016/S0169-4332(01)00730-9Get | $La_{0.7}Ca_{0.3}MnO_3$ | $SrTiO_3$ | 60 |
| H.N.Lee 2007 | 10.1103/PhysRevLett.98.217602 | $PbZr_{0.2}Ti_{0.8}O_3$ | $SrTiO_3$ | 40 |
| S.Venkatesan 2009 | 10.1103/PhysRevB.78.104112 | $TbMnO_3$ | $SrTiO_3$ | 5 |
| Y.Dai 2016 | 10.1063/1.4962853 | $Sr_{0.63}Ba_{0.37}TiO_3$ | $DyScO_3$ | 40 |
| Y.Dai 2016 | 10.1063/1.4962853 | $Sr_{0.875}Ba_{0.125}TiO_3$ | $DyScO_3$ | 18 |
| Y.Dai 2016 | 10.1063/1.4962853 | $SrTiO_3$ | $DyScO_3$ | 11 |
| Y.Dai 2016 | 10.1063/1.4962853 | $SrTiO_3$ | $TbScO_3$ | 9 |
| Y.Dai 2016 | 10.1063/1.4962853 | $SrTiO_3$ | $GdScO_3$ | 5 |
| W.S.Choi 2012 | 10.1021/nl302562f | $LaCoO_3$ | $LaAlO_3$ | 26 |
| S.Zhong 2006 | 10.1557/jmr.2006.0193 | $PbZr_{0.2}Ti_{0.8}O_3$ | $SrTiO_3$ | 12 |
| A.Herpers 2014 | 10.1063/1.4900817 | $Pr_{0.48}Ca_{0.52}MnO_3$ | $SrTiO_3$ | 1.5 |
| D.Fuchs 2002 | 10.1063/1.1461897 | $LaAlO_3$ | LSAT | 3 |
| D.Fuchs 2002 | 10.1063/1.1461897 | $SrTiO_3$ | LSAT | 30 |
| D.Fuchs 2002 | 10.1063/1.1461897 | $La_{0.4}Sr_{0.6}CoO_3$ | LSAT | 90 |
| X.Wang 2012 | 10.1080/14786435.2012.657709 | $BaTiO_3$ | $SrTiO_3$ | 2 |
| X.Wang 2012 | 10.1080/14786435.2012.657709 | $BaTiO_3$ | $SrTiO_3$ | 4 |
| S.Jan 2016 | 10.14279/depositonce-4997 | $NaNbO_3$ | $NdGaO_3$ | 20 |
| S.Jan 2016 | 10.14279/depositonce-4997 | $NaNbO_3$ | $DyScO_3$ | 27 |
| T.Wang 2013 | 10.1063/1.4833248 | $SrTiO_3$ | LSAT | 180 |
| C.M.Foster 1998 | 10.1063/1.360121 | $PbTiO_3$ | $SrTiO_3$ | 150 |
| C.M.Foster 1998 | 10.1063/1.360121 | $PbTiO_3$ | MgO | 10 |
| C.M.Foster 1998 | 10.1063/1.360121 | $PbTiO_3$ | $LaAlO_3$ | 10 |
| S.H.Oh 2004 | 10.1063/1.1690484 | $SrRuO_3$ | $SrTiO_3$ | 10 |
| B.S.Kwak 1992 | 10.1103/PhysRevLett.68.3733 | $PbTiO_3$ | $KTaO_3$ | 34 |
| B.S.Kwak 1992 | 10.1103/PhysRevLett.68.3733 | $PbTiO_3$ | $KTaO_3$ | 250 |
| K.Hirai 2013 | 10.1063/1.4817505 | $SrFeO_{2.5}$ | $DyScO_3$ | 50 |
| R.A.Rao 1999 | 10.1063/1.122749 | $La_{0.8}Ca_{0.2}MnO_3$ | $SrTiO_3$ | 25 |
| R.A.Rao 1999 | 10.1063/1.122749 | $La_{0.8}Ca_{0.2}MnO_3$ | $LaAlO_3$ | 5 |
| L.Peng 2003 | 10.1063/1.1631055 | $SrTiO_3$ | $LaAlO_3$ | 50 |
| L.Peng 2003 | 10.1063/1.1631055 | $SrTiO_3$ | MgO | 50 |
| J.Zhang 2001 | 10.1088/0953-8984/23/33/334211 | $La_{0.05}Ba_{0.95}MnO_3$ | $SrTiO_3$ | 20 |
| G.Catalan 2005 | 10.1103/PhysRevB.72.020102 | $Ba_{0.5}Sr_{0.5}TiO_3$ | MgO | 100 |
| G.Catalan 2005 | 10.1103/PhysRevB.72.020102 | $Ba_{0.5}Sr_{0.5}TiO_3$ | MgO | 200 |
| G.Gao 2007 | 10.1063/1.2429903 | $La_{0.7}Ca_{0.3}MnO_3$ | $SrTiO_3$ | 30 |
| D.H.Kim 2008 | 10.1063/1.2830799 | $BiFeO_3$ | $SrTiO_3$ | 90 |
| T.L.Meyer 2015 | 10.1063/1.4937170 | $La_{1.85}Sr_{0.15}CuO_4$ | $SrTiO_3$ | 15 |
| T.L.Meyer 2015 | 10.1063/1.4937170 | $La_{1.85}Sr_{0.15}CuO_4$ | $LaAlO_3$ | 80 |
| T.L.Meyer 2015 | 10.1063/1.4937170 | $La_{1.85}Sr_{0.15}CuO_4$ | $LaSrAlO_4$ | 35 |
| V.V.Mehta 2015 | 10.1103/PhysRevB.91.144418 | $LaCoO_3$ | $SrTiO_3$ | 15 |
| V.V.Mehta 2015 | 10.1103/PhysRevB.91.144418 | $LaCoO_3$ | $SrTiO_3$ | 73 |
| V.V.Mehta 2015 | 10.1103/PhysRevB.91.144418 | $LaCoO_3$ | LSAT | 15 |
| V.V.Mehta 2015 | 10.1103/PhysRevB.91.144418 | $LaCoO_3$ | LSAT | 73 |
| V.V.Mehta 2015 | 10.1103/PhysRevB.91.144418 | $LaCoO_3$ | $LaAlO_3$ | 8 |
| Y.B.Xu 2016 | 10.1038/srep35172 | $PbTiO_3$ | $LaAlO_3$ | 45 |
| A.I.Khan 2014 | 10.1063/1.4885551 | $PbZr_{0.2}Ti_{0.8}O_3$ | $SrTiO_3$ | 40 |
| E.Breckenfeld 2013 | 10.1039/c3tc31653j | $Sr_{1.04}TiO_3$ | $NdGaO_3$ | 60 |
| E.Breckenfeld 2013 | 10.1039/c3tc31653j | $Sr_{0.96}TiO_3$ | $NdGaO_3$ | 300 |
| S.Gariglio 2007 | 10.1063/1.2740171 | $PbZr_{0.2}Ti_{0.8}O_3$ | $SrTiO_3$ | 25 |
| V.Pena 2006 | 10.1016/j.jpcs.2005.10.022 | $La_{0.67}Ca_{0.33}MnO_3$ | $LaSrAlO_4$ | 5 |
| Xuan Shen 2015 | 10.1063/1.4906430 | $SrZr_{0.68}Ti_{0.32}O_3$ | Ge | 11.4 |
| H.P.Sun 2004 | 10.1063/1.1728300 | $BaTiO_3$ | $SrTiO_3$ | 50 |
| H.P.Sun 2004 | 10.1063/1.1728300 | $BaTiO_3$ | $SrTiO_3$ | 20 |
| K.Daoudi 2010 | 10.1016/j.jallcom.2010.07.035 | $La_{0.7}Sr_{0.3}CoO_3$ | $SrTiO_3$ | 100 |
| M.D. Biegalski 2008 | 10.1063/1.3037216 | $SrTiO_3$ | $DyScO_3$ | 200 |
| J.Santiso 2016 | 10.1021/acsami.6b02896 | $La_{0.7}Sr_{0.3}MnO_3$ | $LaAlO_3$ | 30 |
| A.Petraru 2007 | 10.1063/1.2745277 | $BaTiO_3$ | $SrTiO_3$ | 10 |
| K.Saito 2006 | 10.1143/JJAP.45.7311 | $BiFeO_3$ | $SrTiO_3$ | 90 |
| Z.Fu 2017 | 10.1063/1.4975342 | $BiFeO_3$ | LSAT | 15 |
| H.Terauchi 1992 | 10.1143/JPSJ.61.2194 | $BaTiO_3$ | $SrTiO_3$ | 80 |
| Y.C.Liang 2005 | 10.1016/j.tsf.2005.07.187 | $La_{0.7}Ba_{0.3}MnO_3$ | $SrTiO_3$ | 35 |
| Y.C.Liang 2005 | 10.1016/j.tsf.2005.07.187 | $La_{0.7}Ba_{0.3}MnO_3$ | $SrTiO_3$ | 34 |
| S.Stemmer 1995 | 10.1002/pssa.2211470115 | $PbTiO_3$ | $SrTiO_3$ | 50 |
| S.Stemmer 1995 | 10.1002/pssa.2211470115 | $PbTiO_3$ | $SrTiO_3$ | 100 |
| S.Stemmer 1995 | 10.1002/pssa.2211470115 | $PbTiO_3$ | $SrTiO_3$ | 15 |
| A.Duk 2013 | 10.1063/1.4794405 | $NaNbO_3$ | $TbScO_3$ | 73 |
| Y.Wu 2011 | 10.1063/1.3567297 | $(BiScO_3)_{0.36}(PbTiO_3)_{0.64}$ | $SrTiO_3$ | 15 |
| F.Sandiumenge 2016 | 10.1002/admi.201600106 | $La_{0.7}Sr_{0.3}MnO_3$ | $LaAlO_3$ | 73 |
| F.Sandiumenge 2016 | 10.1002/admi.201600106 | $La_{0.7}Sr_{0.3}MnO_3$ | $LaAlO_3$ | 8 |
| M.Kuroda 2018 | 10.1063/1.5007332 | $SmFeO_3$ | $LaAlO_3$ | 45 |

**TABLE S1.** Original dataset. The dataset includes 82 data on $h_c$ of perovskite oxide thin films from the literature. $h_c$ values are expressed in unit of nm. The highlighted data are the ones used for ML training.

| AWAS | EAAS | ENAS | IEAS | IRAS | OSAS | AWBS | EABS | ENBS | IEBS | IRBS | OSBS | Lattice constant | Lattice mismatch | Poisson ratio | PB factor | Strain energy density factor | Dislocation energy density factor | $h_c$ |
|---|---|---|---|---|---|---|---|---|---|---|---|---|---|---|---|---|---|---|
| 138.9 | 0.47 | 1.1 | 19.18 | 1.36 | 3 | 26.98 | 0.4328 | 1.61 | 28.45 | 0.535 | 3 | 0.3875 | 2.24274 | 0.35 | 0.03709 | 814.844 | -28.6545 | 10 |
| 138.9 | 0.47 | 1.1 | 19.18 | 1.36 | 3 | 26.98 | 0.4328 | 1.61 | 28.45 | 0.535 | 3 | 0.3875 | 2.24274 | 0.35 | 0.03709 | 814.844 | -28.6545 | 30 |
| 138.9 | 0.47 | 1.1 | 19.18 | 1.36 | 3 | 26.98 | 0.4328 | 1.61 | 28.45 | 0.535 | 3 | 0.3875 | 2.24274 | 0.4 | 0.03302 | 915.442 | -28.6545 | 12 |
| 87.62 | 0.05206 | 0.95 | 11.03 | 1.44 | 2 | 47.87 | 0.084 | 1.54 | 43.27 | 0.605 | 4 | 0.396 | 1.40845 | 0.3 | 0.10749 | 147.363 | -14.6732 | 50 |
| 87.62 | 0.05206 | 0.95 | 11.03 | 1.44 | 2 | 47.87 | 0.084 | 1.54 | 43.27 | 0.605 | 4 | 0.396 | 1.40845 | 0.3 | 0.10749 | 147.363 | -14.6732 | 30 |
| 87.62 | 0.05206 | 0.95 | 11.03 | 1.44 | 2 | 47.87 | 0.084 | 1.54 | 43.27 | 0.605 | 4 | 0.396 | 1.40845 | 0.3 | 0.10749 | 147.363 | -14.6732 | 22 |
| 87.62 | 0.05206 | 0.95 | 11.03 | 1.44 | 2 | 47.87 | 0.084 | 1.54 | 43.27 | 0.605 | 4 | 0.3875 | -0.76825 | 0.35 | 0.31612 | 95.6127 | -28.6545 | 100 |
| 87.62 | 0.05206 | 0.95 | 11.03 | 1.44 | 2 | 47.87 | 0.084 | 1.54 | 43.27 | 0.605 | 4 | 0.3875 | -0.76825 | 0.35 | 0.31612 | 95.6127 | -28.6545 | 100 |
| 87.62 | 0.05206 | 0.95 | 11.03 | 1.44 | 2 | 47.87 | 0.084 | 1.54 | 43.27 | 0.605 | 4 | 0.3875 | -0.76825 | 0.35 | 0.31612 | 95.6127 | -28.6545 | 100 |
| 87.62 | 0.05206 | 0.95 | 11.03 | 1.44 | 2 | 47.87 | 0.084 | 1.54 | 43.27 | 0.605 | 4 | 0.379 | -2.94494 | 0.26 | 0.02567 | 1550.54 | -38.6099 | 10 |
| 87.62 | 0.05206 | 0.95 | 11.03 | 1.44 | 2 | 47.87 | 0.084 | 1.54 | 43.27 | 0.605 | 4 | 0.3992 | 2.22791 | 0.27 | 0.04623 | 595.836 | -25.2942 | 12 |
| 87.62 | 0.05206 | 0.95 | 11.03 | 1.44 | 2 | 47.87 | 0.084 | 1.54 | 43.27 | 0.605 | 4 | 0.3992 | 2.22791 | 0.27 | 0.04623 | 595.836 | -25.2942 | 5 |
| 87.62 | 0.05206 | 0.95 | 11.03 | 1.44 | 2 | 47.87 | 0.084 | 1.54 | 43.27 | 0.605 | 4 | 0.3992 | 2.22791 | 0.27 | 0.04623 | 595.836 | -25.2942 | 10 |
| 87.62 | 0.05206 | 0.95 | 11.03 | 1.44 | 2 | 47.87 | 0.084 | 1.54 | 43.27 | 0.605 | 4 | 0.3992 | 2.22791 | 0.27 | 0.04623 | 595.836 | -25.2942 | 3.2 |
| 87.62 | 0.05206 | 0.95 | 11.03 | 1.44 | 2 | 47.87 | 0.084 | 1.54 | 43.27 | 0.605 | 4 | 0.3992 | 2.22791 | 0.27 | 0.04623 | 595.836 | -25.2942 | 5 |
| 87.62 | 0.05206 | 0.95 | 11.03 | 1.44 | 2 | 47.87 | 0.084 | 1.54 | 43.27 | 0.605 | 4 | 0.3992 | 2.22791 | 0.27 | 0.04623 | 595.836 | -25.2942 | 10 |
| 87.62 | 0.05206 | 0.95 | 11.03 | 1.44 | 2 | 47.87 | 0.084 | 1.54 | 43.27 | 0.605 | 4 | 0.3992 | 2.22791 | 0.27 | 0.04623 | 595.836 | -25.2942 | 10 |
| 87.62 | 0.05206 | 0.95 | 11.03 | 1.44 | 2 | 47.87 | 0.084 | 1.54 | 43.27 | 0.605 | 4 | 0.3992 | 2.22791 | 0.27 | 0.04623 | 595.836 | -25.2942 | 10 |
| 87.62 | 0.05206 | 0.95 | 11.03 | 1.44 | 2 | 47.87 | 0.084 | 1.54 | 43.27 | 0.605 | 4 | 0.3992 | 2.22791 | 0.27 | 0.04623 | 595.836 | -25.2942 | 15 |
| 87.62 | 0.05206 | 0.95 | 11.03 | 1.44 | 2 | 47.87 | 0.084 | 1.54 | 43.27 | 0.605 | 4 | 0.396 | 1.40845 | 0.3 | 0.10749 | 147.363 | -14.6732 | 15 |
| 87.62 | 0.05206 | 0.95 | 11.03 | 1.44 | 2 | 47.87 | 0.084 | 1.54 | 43.27 | 0.605 | 4 | 0.3992 | 2.22791 | 0.27 | 0.04623 | 595.836 | -25.2942 | 10 |
| 96.85 | 0.12724 | 0.977 | 12.5 | 1.54 | 2.18 | 90.09 | 0.3869 | 1.5649 | 29.9 | 0.594 | 3.41 | 0.383 | -0.98242 | 0.26 | 0.23306 | 80.4425 | -17.993 | 42 |
| 96.85 | 0.12724 | 0.977 | 12.5 | 1.54 | 2.18 | 90.09 | 0.3869 | 1.5649 | 29.9 | 0.594 | 3.41 | 0.383 | -0.98242 | 0.26 | 0.23306 | 80.4425 | -17.993 | 80 |

| Author/year | Thin film | Substrate | AWAF | EAAF | ENAF | IEAF | IRAF | OSAF | AWBF | EABF | ENBF | IEBF | IRBF | OSBF |
|---|---|---|---|---|---|---|---|---|---|---|---|---|---|---|
| H.L.Ju 1998 | $La_{0.66}Sr_{0.33}MnO_3$ | $LaAlO_3$ | 120.589 | 0.32738 | 1.0395 | 16.2987 | 1.3728 | 2.64 | 54.94 | 0 | 1.55 | 39.1182 | 0.5577 | 3.3 |
| H.L.Ju 1998 | $La_{0.66}Sr_{0.33}MnO_3$ | $LaAlO_3$ | 120.589 | 0.32738 | 1.0395 | 16.2987 | 1.3728 | 2.64 | 54.94 | 0 | 1.55 | 39.1182 | 0.5577 | 3.3 |
| A.Tebano 2006 | $La_{0.7}Sr_{0.3}MnO_3$ | $LaAlO_3$ | 123.516 | 0.34462 | 1.055 | 16.735 | 1.384 | 2.7 | 54.94 | 0 | 1.55 | 38.929 | 0.565 | 3.3 |
| S.H.Lim 2007 | $BiFeO_3$ | $SrTiO_3$ | 209 | 0.9424 | 2.02 | 25.56 | 1.17 | 3 | 55.85 | 0.151 | 1.96 | 30.65 | 0.55 | 3 |
| Y.H.Chu 2007 | $BiFeO_3$ | $SrTiO_3$ | 209 | 0.9424 | 2.02 | 25.56 | 1.17 | 3 | 55.85 | 0.151 | 1.96 | 30.65 | 0.55 | 3 |
| S.Geprags 2011 | $BiFeO_3$ | $SrTiO_3$ | 209 | 0.9424 | 2.02 | 25.56 | 1.17 | 3 | 55.85 | 0.151 | 1.96 | 30.65 | 0.55 | 3 |
| J.L.Maurice 2003 | $La_{0.66}Sr_{0.33}MnO_3$ | $SrTiO_3$ | 120.589 | 0.32738 | 1.0395 | 16.2987 | 1.3728 | 2.64 | 54.94 | 0 | 1.55 | 39.1182 | 0.5577 | 3.3 |
| P.M.Vaghefi 2017 | $La_{0.7}Sr_{0.3}MnO_3$ | $SrTiO_3$ | 123.516 | 0.34462 | 1.055 | 16.735 | 1.384 | 2.7 | 54.94 | 0 | 1.55 | 38.929 | 0.565 | 3.3 |
| L. Ranno 2002 | $La_{0.7}Sr_{0.3}MnO_3$ | $SrTiO_3$ | 123.516 | 0.34462 | 1.055 | 16.735 | 1.384 | 2.7 | 54.94 | 0 | 1.55 | 38.929 | 0.565 | 3.3 |
| L.Qiao 2011 | $LaAlO_3$ | $SrTiO_3$ | 138.9 | 0.47 | 1.1 | 19.18 | 1.36 | 3 | 26.98 | 0.4328 | 1.61 | 28.45 | 0.535 | 3 |
| T.Suzuki 1999 | $BaTiO_3$ | $SrTiO_3$ | 137.3 | 0.1446 | 0.89 | 10 | 1.61 | 2 | 47.87 | 0.084 | 1.54 | 43.27 | 0.605 | 4 |
| A.Visinoiu 2002 | $BaTiO_3$ | $SrTiO_3$ | 137.3 | 0.1446 | 0.89 | 10 | 1.61 | 2 | 47.87 | 0.084 | 1.54 | 43.27 | 0.605 | 4 |
| R.Guo 2016 | $BaTiO_3$ | $SrTiO_3$ | 137.3 | 0.1446 | 0.89 | 10 | 1.61 | 2 | 47.87 | 0.084 | 1.54 | 43.27 | 0.605 | 4 |
| J.Zhu 2006 | $BaTiO_3$ | $SrTiO_3$ | 137.3 | 0.1446 | 0.89 | 10 | 1.61 | 2 | 47.87 | 0.084 | 1.54 | 43.27 | 0.605 | 4 |
| M.Fujimoto 2002 | $BaTiO_3$ | $SrTiO_3$ | 137.3 | 0.1446 | 0.89 | 10 | 1.61 | 2 | 47.87 | 0.084 | 1.54 | 43.27 | 0.605 | 4 |
| M.Fujimoto 2002 | $BaTiO_3$ | $SrTiO_3$ | 137.3 | 0.1446 | 0.89 | 10 | 1.61 | 2 | 47.87 | 0.084 | 1.54 | 43.27 | 0.605 | 4 |
| G.H.Lee 2001 | $BaTiO_3$ | $SrTiO_3$ | 137.3 | 0.1446 | 0.89 | 10 | 1.61 | 2 | 47.87 | 0.084 | 1.54 | 43.27 | 0.605 | 4 |
| H.I.Seo 2020 | $BaTiO_3$ | $SrTiO_3$ | 137.3 | 0.1446 | 0.89 | 10 | 1.61 | 2 | 47.87 | 0.084 | 1.54 | 43.27 | 0.605 | 4 |
| H.I.Seo 2020 | $BaTiO_3$ | $SrTiO_3$ | 137.3 | 0.1446 | 0.89 | 10 | 1.61 | 2 | 47.87 | 0.084 | 1.54 | 43.27 | 0.605 | 4 |
| H.J.Lee 2016 | $BiFeO_3$ | $SrTiO_3$ | 209 | 0.9424 | 2.02 | 25.56 | 1.17 | 3 | 55.85 | 0.151 | 1.96 | 30.65 | 0.55 | 3 |
| Y.S.Kim 2005 | $BaTiO_3$ | $SrTiO_3$ | 137.3 | 0.1446 | 0.89 | 10 | 1.61 | 2 | 47.87 | 0.084 | 1.54 | 43.27 | 0.605 | 4 |
| D.Zhang 2019 | $SrCoO_3$ | LSAT | 87.62 | 0.05206 | 0.95 | 11.03 | 1.44 | 2 | 58.93 | 0.6633 | 1.88 | 51.3 | 0.53 | 4 |
| D.Zhang 2019 | $SrCoO_3$ | LSAT | 87.62 | 0.05206 | 0.95 | 11.03 | 1.44 | 2 | 58.93 | 0.6633 | 1.88 | 51.3 | 0.53 | 4 |

**TABLE S2.** Final dataset. The refined dataset with 23 data on $h_c$ of perovskite oxide thin films used for ML training. Ionic features, phase features, and PB model features are also shown. Lattice constant and PB factor are expressed in unit of nm. Strain energy density factor and dislocation energy density factor are expressed in units of GPa and GPa·nm respectively. Ionic features are presented using the following notation: AW (atomic weight), EA, (electron affinity), EN (electronegativity), IE (ionization energy), IR (ionic radius), OS (oxidation states), of A or B cation in film (F) or substrate (S) material of perovskite $ABO_3$.